# Mechanical action of inhomogeneously polarized optical fields and detection of the internal energy flows

**Aleksandr Bekshaev and Svetlana Sviridova**

I.I. Mechnikov National University, Dvorianska 2, Odessa, 65082, Ukraine
E-mail: bekshaev@onu.edu.ua

**Abstract**

We analyze numerically the correspondence between the mechanical action, experienced by a spherical microparticle, and the internal energy flows as well as spatial and polarization inhomogeneity of the light field incident on the particle. The inhomogeneous incident field is modelled by superposition of two plane waves, the mechanical action is calculated via the Mie theory for dielectric and conducting particles of different sizes and optical properties. It is shown that both spin and orbital components of the field momentum can produce the mechanical action, which can thus be used for experimental study of the internal energy flows in light beams. However, exact value and even direction of the force applied to a particle depends on many details of the field-particle interaction. Besides, forces that are not associated with any sort of the energy flow (we attribute them to the gradient force owing to the inhomogeneous intensity and to the dipole force emerging due to inhomogeneous polarization) can strongly modify the observed mechanical action. The results should be taken into account in experiments employing the motion of probing particles suspended within optical fields. We suppose that situations where an optical field exerts the polarization-dependent mechanical action on particles can be treated as a new form of the spin-orbit interaction of light.

### 1. Introduction

During the past years, internal energy flows in light fields (optical currents) are studied with growing interest [1–19]. They attract particular attention as physical characteristics of light beams with clear and unambiguous physical meaning, valid both for scalar and vector beams with arbitrary polarization properties [1−9]. Specific patterns of the energy circulation, appearing in connection to singular points of the optical fields, stipulate intensive investigation of optical currents within the frame of singular optics [1−5,7−9,15−17]. The internal flows provide a convenient means for characterization of spatial beam transformation during the free propagation as well as in presence of obstacles [5,6,13,18] and constitute a suitable set of the optical field parameters, immediately oriented at rapidly developing applications related to optical trapping, sorting and micromanipulation [20].

In case of free-space monochromatic electromagnetic fields with radiation frequency $\omega$, to which we are restricted in this paper, the electric and magnetic vectors can be taken in forms $\mathrm{Re}\left[\mathbf{E}\exp(-i\omega t)\right]$, $\mathrm{Re}\left[\mathbf{H}\exp(-i\omega t)\right]$. The complex vector amplitudes $\mathbf{E}$ and $\mathbf{H}$ provide suitable

representations of the time-average energy parameters of such fields [21]. The energy density equals to

$$w = \frac{g}{2}\left(|\mathbf{E}|^2 + |\mathbf{H}|^2\right), \qquad (1)$$

and the energy flow density is expressed by the Poynting vector **S** or electromagnetic momentum density **p** distributions:

$$\mathbf{S} = c^2 \mathbf{p} = gc\,\mathrm{Re}\left(\mathbf{E}^* \times \mathbf{H}\right) \qquad (2)$$

( $g = (8\pi)^{-1}$ in the Gaussian system of units, $c$ is the light velocity). Since quantities **S** and **p** are proportional, in many cases they can be considered as equivalent so the energy flow pattern can be characterized by the field momentum distribution and vice versa. The total energy flow (2) can be subdivided into the spin and orbital parts, $\mathbf{p} = \mathbf{p}_S + \mathbf{p}_O$, according to which sort of the beam angular momentum they are able to generate [9,10,14]:

$$\mathbf{p}_S = \frac{g}{4\omega}\mathrm{Im}\left[\nabla \times \left(\mathbf{E}^* \times \mathbf{E} + \mathbf{H}^* \times \mathbf{H}\right)\right], \quad \mathbf{p}_O = \frac{g}{2\omega}\mathrm{Im}\left[\mathbf{E}^* \cdot (\nabla)\mathbf{E} + \mathbf{H}^* \cdot (\nabla)\mathbf{H}\right]. \qquad (3)$$

The spin flow is usually associated with inhomogeneous circular polarization while the orbital one owes to the explicit energy redistribution within the optical beam. Peculiar properties of the spin and orbital contributions reflect specific features of the macroscopic energy transfer ($\mathbf{p}_O$) and "intrinsic" rotation associated with the spin of photons ($\mathbf{p}_S$). Quantities (3) provide deeper insight into thin details of the light field evolution and allow to describe mutual conversion of the light energy between the spin and orbital degrees of freedom [17,18].

Nevertheless, in spite of the above-listed attractive features, wide usage of the internal flow parameters is hampered by the lack of direct methods for their measurement [9,17,22]. As a promising approach to their detection and quantification, the motion of small probing particles suspended within the optical field was proposed and experimentally tested [14,17,23]. It is based on assumption that the force acting on a particle is proportional to the local value of the field momentum. However, its correctness is rather questionable. Recent calculations performed for various models of particle and particle-field interaction [12,14,20] has shown that the force applied to a particle is, of course, related to the field momentum but in rather intricate ways and the simple proportional dependence occurs likely as an exclusion. Even the physical model of the momentum transfer from the field to a particle is not clear. For example, usual explanation of the simple situation when a particle absorbs some part of the light energy and takes over the corresponding momentum, associated with this energy, is not applicable to the spin momentum of a circularly polarized wave. Such a wave, as well as any its fragment, carries the "pure" angular momentum that can cause the spinning motion of the absorbing particle, but there is no clear understanding whether and how the translational or orbital motion will appear in this situation [12,17]. What is more, any particle placed in the electromagnetic field distorts it, sometimes very strongly [24], so the real field acting on the particle has little in common with the "original" free-of-particle field whose parameters are the main subjects of interest.

In this paper, we present an attempt of direct calculation of relations between the force applied to the particle and the energy flow in the optical field that existed before the particle is placed there ("incident field" with vector amplitudes **E** and **H**). The main idea is to determine the electromagnetic field disturbed by the presence of a particle, to calculate its momentum and to compare the result with the initial momentum carried by the "pure" incident field.

Due to the particle presence, the scattered field $\mathbf{E}_{sc}$, $\mathbf{H}_{sc}$ emerges that should be added to the incident field **E**, **H** [24] so the total field momentum density is changed by

$$\Delta\mathbf{p} = \frac{g}{c}\mathrm{Re}\left[\left(\mathbf{E}^* + \mathbf{E}^*_{sc}\right) \times \left(\mathbf{H} + \mathbf{H}_{sc}\right) - \mathbf{E}^* \times \mathbf{H}\right] = \frac{g}{c}\mathrm{Re}\left(\mathbf{E}^*_{sc} \times \mathbf{H}_{sc} + \mathbf{E}^* \times \mathbf{H}_{sc} + \mathbf{E}^*_{sc} \times \mathbf{H}\right). \qquad (4)$$

The change of the field momentum results in the recoil force applied to the particle; the force can be determined by the field momentum flux through the spherical surface $A_R$ with radius $R \to \infty$ surrounding the particle:

$$\mathbf{F} = -c\oint_{A_R} \Delta \mathbf{p}\, dA = -cR^2 \oint \Delta \mathbf{p}\, d\Omega \qquad (5)$$

($d\Omega$ means integration over the solid angle). This force depends on the particle position and our aim is to find correspondence between $\mathbf{F}$ and the incident field momentum in the point where the particle is placed.

In principle, the light scattered by a spherical particle can be rather simply calculated following the Mie theory [24]. However, the original Mie approach is only applicable to an incident field in the form of a plane monochromatic wave whereas we are interested in studying the incident fields with inhomogeneous configuration and well developed pattern of the internal energy flows. Such field configurations can be modeled by superposition of many plane waves and then the scattered light can be found as a sum of the Mie results obtained for every member of the superposition. This direct and conceptually simple mode of operation leads, though, to very extensive computing because even for a single plane wave the Mie theory calculations deal with slowly converging series of quickly oscillating functions and are generally cumbersome. To avoid the unnecessary complications that, additionally, may obscure the physical picture, in this paper we consider the simplest models of spatially inhomogeneous optical fields consisting of only two plane waves, which, nevertheless, adequately represent the physical nature of the spin and orbital flows in real inhomogeneous optical beams.

## 2. Model description

### 2.1. Incident field configuration

Geometrical conditions of the problem are illustrated by Fig. 1. The center of a spherical particle is situated in the origin of the laboratory frame ($xyz$) and is illuminated by the light coming from the lower hemisphere ($z < 0$). Contrary to the usual geometry [24], we intend to consider the case of inhomogeneous incident field distribution over the nominal transverse plane $z = 0$. Such a field can be represented via superposition of plane waves differently oriented with respect to the nominal longitudinal axis $z$. The $j$-th plane wave propagates along axis $z_j$ that deviates from the laboratory axis $z$ by the incidence angle $\gamma_j$; in what follows, we restrict ourselves to the case when angles $\gamma_j$ lie in the coordinate plane ($yz$). Further, we introduce an "own" coordinate frame ($x\, y_j\, z_j$) associated with each member of the plane-wave superposition; the "own" and the laboratory coordinates are united by known relations

$$z_j = z\cos\gamma_j + y\sin\gamma_j, \quad y_j = -z\sin\gamma_j + y\cos\gamma_j. \qquad (6)$$

In its own frame, the electric and magnetic fields of a separate plane-wave component are described by equations

$$\mathbf{E}_{aj}(x, y_j, z_j) = \mathbf{E}_j \exp(ikz_j) = \begin{pmatrix} E_{xj} \\ E_{xj} \end{pmatrix} \exp(ikz_j), \quad \mathbf{H}_{aj}(x, y_j, z_j) \equiv \mathbf{H}_j \exp(ikz_j) = \mathbf{e}_j \times \mathbf{E}_{aj}(z_j) \qquad (7)$$

where $E_{jx}$ and $E_{jy}$ are constants, $\mathbf{e}_j$ is the unit vector of the $z_j$-axis and $k$ is the wave number of the incident radiation. The optical field, created by wave (7) in the common reference plane $z = 0$, is generally inhomogeneous and in the laboratory coordinates can be written in the form

$$\mathbf{E}_{aj}(x, y) = \begin{pmatrix} E_{xj} \\ E_{yj}\cos\gamma_j \\ -E_{yj}\sin\gamma_j \end{pmatrix} \exp(ikz_j), \quad \mathbf{H}_{aj}(x, y) = \begin{pmatrix} -E_{yj} \\ E_{xj}\cos\gamma_j \\ -E_{xj}\sin\gamma_j \end{pmatrix} \exp(ikz_j) \qquad (8)$$

where $z_j$ is related to $y$ and $z$ by the first Eq. (6).

For the simplest superposition consisting of only two plane waves, the electric and magnetic strengths of the incident optical field equal to

$$\mathbf{E} = \mathbf{E}_{a1} + \mathbf{E}_{a2}, \quad \mathbf{H} = \mathbf{H}_{a1} + \mathbf{H}_{a2}. \qquad (9)$$

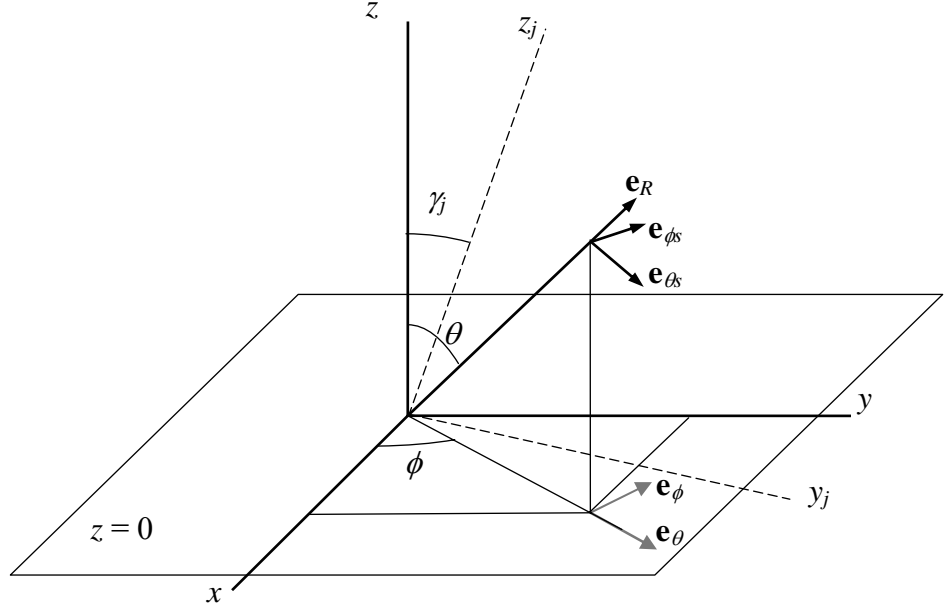

Fig. 1. Geometrical conditions of the light scattering analysis. The particle is situated in the coordinate origin, incident light comes from the lower hemisphere; other explanations see in text.

After some algebra we find the optical field energy density (1) and components of the spin and orbital momentum density (3) in the following representations:

$$w = g\left[|\mathbf{E}_1|^2 + |\mathbf{E}_2|^2 + \cos^2\frac{\gamma_1 - \gamma_2}{2} D(y,z)\right]; \qquad (10)$$

$$p_{Sx} = \frac{g}{2c}\sin(\gamma_1 - \gamma_2)\left[\left(E^*_{x2}E_{y1} - E^*_{y2}E_{x1}\right)e^{ik(z_1 - z_2)} + \left(E^*_{y1}E_{x2} - E^*_{x1}E_{y2}\right)e^{ik(z_2 - z_1)}\right], \qquad (11)$$

$$p_{Sy} = \frac{g}{2c}\sin^2\frac{\gamma_1 - \gamma_2}{2}(\sin\gamma_1 + \sin\gamma_2)D(y,z), \qquad (12)$$

$$p_{Sz} = \frac{g}{2c}\sin^2\frac{\gamma_1 - \gamma_2}{2}(\cos\gamma_1 + \cos\gamma_2)D(y,z); \qquad (13)$$

$$p_{Ox} = 0, \qquad (14)$$

$$p_{Oy} = \frac{g}{c}\left[|\mathbf{E}_1|^2\sin\gamma_1 + |\mathbf{E}_2|^2\sin\gamma_2 + \frac{1}{2}\cos^2\frac{\gamma_1 - \gamma_2}{2}(\sin\gamma_1 + \sin\gamma_2)D(y,z)\right], \qquad (15)$$

$$p_{Oz} = \frac{g}{c}\left[|\mathbf{E}_1|^2\cos\gamma_1 + |\mathbf{E}_2|^2\cos\gamma_2 + \frac{1}{2}\cos^2\frac{\gamma_1 - \gamma_2}{2}(\cos\gamma_1 + \cos\gamma_2)D(y,z)\right] \qquad (16)$$

where

$$D(y,z) = \left(E^*_{x2}E_{x1} + E^*_{y2}E_{y1}\right)e^{ik(z_1 - z_2)} + \left(E^*_{x1}E_{x2} + E^*_{y1}E_{y2}\right)e^{ik(z_2 - z_1)}.$$

Eqs. (11) – (16) show that our simple superposition can serve a model of real inhomogeneous fields with non-zero spin and orbital flows. The *y*- and *z*-components of the orbital flow contain trivial contributions owing to the longitudinal energy transfer by both superposed plane waves (first and second summands in brackets of (15) and (16)); the internal flows 'per se' are expressed by the last ("interference") summands depending on the incident field inhomogeneity. Noteworthy (Eq. (14)), in the considered field geometry, the $x$-component of the orbital flow is absent and the whole $x$-directed flow is of the spin nature (11). We can also account for the particle displacement from the coordinate origin by equivalent shift of the inhomogeneous field pattern: if, say, the initial phase of the second wave changes, $\mathbf{E}_2 \to \mathbf{E}_2 e^{i\delta}$, it is equivalent to the particle position at

$$y = -\frac{\delta}{k(\sin\gamma_1 - \sin\gamma_2)}.$$

2.2. Scattered field and mechanical action

The light scattered by the spherical particle illuminated by a plane monochromatic wave can be calculated with using the Mie theory [24]. To find the field mechanical action (5), one should know the scattered field at $R \to \infty$. For such conditions, the scattered field produced by the $j$-th plane wave (8) can be found via relations

$$\mathbf{E}_{scj} = \frac{e^{ikR}}{-ikR}\mathbf{E}_{sj}, \quad \mathbf{H}_{scj} = \frac{e^{ikR}}{-ikR}\mathbf{H}_{sj} \tag{17}$$

where

$$\mathbf{E}_{sj} = \begin{pmatrix} E_{\theta sj} \\ E_{\phi sj} \end{pmatrix} = \begin{pmatrix} S_2 & 0 \\ 0 & S_1 \end{pmatrix}\begin{pmatrix} E_{\theta j} \\ E_{\phi j} \end{pmatrix} = \begin{pmatrix} S_2 & 0 \\ 0 & S_1 \end{pmatrix}\begin{pmatrix} E_{xj}\cos\phi_j + E_{yj}\sin\phi_j \\ -E_{xj}\sin\phi_j + E_{yj}\cos\phi_j \end{pmatrix},$$

$$\mathbf{H}_{sj} = \begin{pmatrix} H_{\theta sj} \\ H_{\phi sj} \end{pmatrix} = \begin{pmatrix} 0 & -1 \\ 1 & 0 \end{pmatrix}\begin{pmatrix} E_{\theta sj} \\ E_{\phi sj} \end{pmatrix}, \tag{18}$$

$S_1 \equiv S_1(\cos\theta_j)$ and $S_2 \equiv (\cos\theta_j)$ are elements of the scattering matrix [24] depending on the wave number $k$, particle radius $a$ and the complex refraction index $m$, and Cartesian and spherical coordinates are measured in the frame $(x, y_j, z_j)$ associated with the $j$-th incident plane wave (see Fig. 1). The scattered field is completely transverse: all the components of Eq. (18) are orthogonal to the unit vector $\mathbf{e}_R$. In the simplest case of the Rayleigh scattering, when the particle is much less than the wavelength,

$$S_1 = -i(ka)^3\frac{m^2-1}{m^2+2}, \quad S_2 = S_1\cos\theta_j, \tag{19}$$

in more general situations, $S_1$ and $S_2$ are expressed via the spherical functions [24] and can be calculated numerically. Each plane wave of the incident field is scattered independently; so the resulting scattered field can be found by the vector summation of the results obtained via Eqs. (18). In view of relations (17) and for future convenience, we represent it in the form

$$\mathbf{E}_{sc1} + \mathbf{E}_{sc2} = \mathbf{E}_s\frac{e^{ikR}}{kR}, \quad \mathbf{H}_{sc1} + \mathbf{H}_{sc2} = \mathbf{H}_s\frac{e^{ikR}}{kR}. \tag{20}$$

Introduced quantities $\mathbf{E}_s$ and $\mathbf{H}_s$ are merely the scattered field amplitudes without the "spherical wave" factor $e^{ikR}/(-ikR)$.

Now, by using Eqs. (7), Eq. (4) can be written as

$$\Delta\mathbf{p} = \frac{g}{ckR}\mathrm{Re}\left[\frac{\mathbf{E}_s^*\times\mathbf{H}_s}{kR} + \mathbf{E}_1^*\times\mathbf{H}_s e^{ik(R-z_1)} + \mathbf{E}_2^*\times\mathbf{H}_s e^{ik(R-z_2)} + \mathbf{E}_s^*\times\mathbf{H}_1 e^{-ik(R-z_1)} + \mathbf{E}_s^*\times\mathbf{H}_2 e^{-ik(R-z_2)}\right] \tag{21}$$

which should be substituted into (3). Then, due to Eq. (18), the first term in brackets transforms to

$$\mathbf{F}_s = -\frac{g}{(kR)^2}\oint_{A_R}\mathbf{E}_s^*\times\mathbf{H}_s dA = -\frac{g}{k^2}\int_0^{2\pi}d\phi\int_0^\pi\left(|E_{\theta s}|^2 + |E_{\phi s}|^2\right)\mathbf{e}_R(\theta,\phi)\sin\theta d\theta \tag{22}$$

and can be evaluated numerically. Note that for other terms, numerical integration is practically impossible because of quickly oscillating interference factors $\exp[\pm ik(R-z_j)]$. However, their behavior at $R \to \infty$ can be evaluated analytically. They contain integrals in the form with easily derived asymptotic:

$$\int_0^\pi F(\theta) e^{\pm ikR(1-\cos\theta)} \sin\theta\, d\theta = \frac{e^{\pm ikR}}{\mp ikR}\left[e^{\pm ikR} F(\theta)_{\theta=\pi} - e^{\mp ikR} F(\theta)_{\theta=0}\right] + O\left(\frac{1}{k^2 R^2}\right);$$

the latter transformation is valid for any function $F(\theta)$ with sufficiently regular behavior. In application to summands of (21) this rule yields

$$R^2 \operatorname{Re} \int_0^{2\pi} d\phi_j \int_0^\pi \left[\mathbf{E}_j^* \times \mathbf{H}_s e^{ik(R-z_j)} + \mathbf{E}_s^* \times \mathbf{H}_j e^{-ik(R-z_j)}\right] \sin\theta_j\, d\theta_j$$

$$= R^2 \frac{2\pi}{kR} \operatorname{Im}\left[e^{2ikR}\left(\mathbf{E}_j^* \times \mathbf{H}_s\right)_{\theta_j=\pi} - \left(\mathbf{E}_j^* \times \mathbf{H}_s\right)_{\theta_j=0} - e^{-2ikR}\left(\mathbf{E}_s^* \times \mathbf{H}_j\right)_{\theta_j=\pi} + \left(\mathbf{E}_s^* \times \mathbf{H}_j\right)_{\theta_j=0}\right]. \quad (23)$$

Further, since $\mathbf{H}_j = \mathbf{e}_{zj} \times \mathbf{E}_j$ and $\mathbf{H}_s = \mathbf{e}_R \times \mathbf{E}_s$,

$$\mathbf{E}_s^* \times \mathbf{H}_j = \mathbf{E}_s^* \times \left(\mathbf{e}_{zj} \times \mathbf{E}_j\right) = \mathbf{e}_{zj}\left(\mathbf{E}_s^* \cdot \mathbf{E}_j\right) - \mathbf{E}_j\left(\mathbf{e}_{zj} \cdot \mathbf{E}_s^*\right),$$

$$\mathbf{E}_j^* \times \mathbf{H}_s = \mathbf{E}_j^* \times \left(\mathbf{e}_R \times \mathbf{E}_s\right) = \mathbf{e}_R\left(\mathbf{E}_j^* \cdot \mathbf{E}_s\right) - \mathbf{E}_s\left(\mathbf{e}_R \cdot \mathbf{E}_j^*\right).$$

In point $\theta_j = 0$ $\mathbf{e}_R = \mathbf{e}_{zj}$, in point $\theta_j = \pi$ $\mathbf{e}_R = -\mathbf{e}_{zj}$. Accordingly, second summands in the above equations vanish and

$$\left(\mathbf{E}_j^* \times \mathbf{H}_s\right)_{\theta_j=0} = \left(\mathbf{E}_s^* \times \mathbf{H}_j\right)_{\theta_j=0}^*, \quad \left(\mathbf{E}_s^* \times \mathbf{H}_j\right)_{\theta_j=\pi} = -\left(\mathbf{E}_j^* \times \mathbf{H}_s\right)_{\theta_j=\pi}^*. \quad (24)$$

As a result, contribution of points $\theta_j = \pi$ in expression (23) vanishes and, combining (23), (24), (21) and (3), one obtains

$$\mathbf{F} = \mathbf{F}_s - g\frac{4\pi}{k^2} \operatorname{Im}\left[\left(\mathbf{E}_s^* \times \mathbf{H}_1\right)_{\theta_1=0} + \left(\mathbf{E}_s^* \times \mathbf{H}_2\right)_{\theta_2=0}\right] \quad (25)$$

where $\mathbf{F}_s$ is given by Eq. (22), $\mathbf{E}_s$ is determined by Eqs. (20), (17) and (18), and $\mathbf{H}_j$ is the amplitude of the incident plane-wave component as defined in Eq. (7). Note that due to the accepted incident field geometry (Fig. 1), both summands in brackets of (25) are vectors belonging to plane ($yz$), and the $x$-component of force $F_x = F_{xs}$ is fully determined by the momentum of the scattered field alone (22).

So the procedure of calculating the mechanical action of the field described in Sec. 1.1 is clear and can be realized numerically.

### 3. Simple symmetrical configurations of the incident field

Eqs. (8) and (10) – (16) allow to analyze various situations. Our purpose of studying the role of spin and orbital flows can be achieved via considering some symmetric superpositions of plane waves (8) that appear if

$$\gamma_1 = -\gamma_2 = \gamma \quad (26)$$

and both waves are identical with possible phase shift, i.e.

$$E_{x2} = E_{x1} e^{i\delta}, \quad E_{y2} = E_{y1} e^{i\delta}. \quad (27)$$

Then

$$\mathbf{E} = 2\exp\left(ikz\cos\gamma + i\frac{\delta}{2}\right)\left(\mathbf{e}_x E_{x1} \cos\Phi + \mathbf{e}_y E_{y1} \cos\gamma \cos\Phi - \mathbf{e}_z iE_{y1} \sin\gamma \sin\Phi\right), \quad (28)$$

$$\mathbf{H} = 2\exp\left(ikz\cos\gamma + i\frac{\delta}{2}\right)\left(-\mathbf{e}_x E_{y1} \cos\Phi + \mathbf{e}_y E_{x1} \cos\gamma \cos\Phi - \mathbf{e}_z iE_{x1} \sin\gamma \sin\Phi\right) \quad (29)$$

where

$$\Phi = ky\sin\gamma - \frac{\delta}{2} \quad (30)$$

and Eqs. (10) – (16) reduce to

$$w = 2g\left(|E_{x1}|^2 + |E_{y1}|^2\right)\left(1 + \cos^2\gamma \cos 2\Phi\right), \tag{31}$$

$$p_{Sx} = -\frac{g}{c}i\left(E_{x1}E_{y1}^* - E_{y1}E_{x1}^*\right)\sin 2\gamma \sin 2\Phi, \tag{32}$$

$$p_{Sy} = p_{Oy} = p_{Ox} = 0,$$

$$p_{Sz} = \frac{2g}{c}\left(|E_{x1}|^2 + |E_{y1}|^2\right)\cos\gamma \sin^2\gamma \cos 2\Phi, \tag{33}$$

$$p_{Oz} = \frac{2g}{c}\left(|E_{x1}|^2 + |E_{y1}|^2\right)\cos\gamma\left(1 + \cos^2\gamma \cos 2\Phi\right). \tag{34}$$

Noticeably, the spin momentum (32) agrees with $z$-component of the "paraxial" relation [9,17]

$$\mathbf{p}_S = -\frac{1}{2\omega c}\left[\mathbf{e}_z \times \nabla_\perp s_3\right] \tag{35}$$

where

$$s_3 = icg\left(E_x E_y^* - E_x^* E_y\right) = 2icg\left(E_{x1}E_{y1}^* - E_{x1}^* E_{y1}\right)\left(1 + \cos 2\Phi\right) \tag{36}$$

is the "third Stokes parameter" describing the polarization ellipticity in the plane ($xy$) [9].

An interesting consequence of the above results is that the spin flow does not vanish even in case when the incident field polarization is linear and everywhere the same – contrary to usual notions, along which the spin flow is associated with the inhomogeneous circular or elliptic polarization [9,10,12]. For example, in $x$- or $y$-polarized fields where

$$E_{x1} = 0, E_{y1} \neq 0 \quad \text{or} \quad E_{y1} = 0, E_{x1} \neq 0, \tag{37}$$

$x$-component (32) of the spin momentum behaves "normally", $p_{Sx} = 0$, but its $z$-component (33) is still non-zero. This paradoxical feature can be explained if the longitudinal field components are taken into account. In general, in planes ($xz$) and ($yz$), the field of Eqs. (28) and (29) is elliptically polarized and the presence of circular component can be characterized by corresponding analogs $s_3^{(xz)}$ and $s_3^{(yz)}$ of the third Stokes parameter (36). They can be calculated similarly to (36); the only precaution is that, because of the field non-paraxiality, contributions of electric and magnetic fields generally differ and must therefore be added with equal weights:

$$s_3^{(xz)} = \frac{i}{2}cg\left(E_x E_z^* - E_x^* E_z + H_x H_z^* - H_x^* H_z\right) = 0,$$

$$s_3^{(yz)} = \frac{i}{2}cg\left(E_y E_z^* - E_y^* E_z + H_y H_z^* - H_y^* H_z\right) = -gc\left(|E_{x1}|^2 + |E_{y1}|^2\right)\sin 2\gamma \sin 2\Phi \tag{38}$$

(in plane ($xz$), projections of the electric and magnetic vectors rotate oppositely while in plane ($yz$) they rotate identically). Interestingly, relation between the spin momentum (33) and parameter (38),

$$p_{Sz} = -\frac{1}{2\omega c}\frac{\partial s_3^{(yz)}}{\partial y}, \tag{39}$$

exactly reproduces the paraxial formula (35) if the "transverse" ($xy$) plane is replaced by the ($yz$) plane and the "longitudinal" direction is associated with $\mathbf{e}_x$. Usual rules for the spin flow, derived for paraxial beams [9,10], are also applicable. For example, within the ($yz$) plane, the spin flow is oriented along the constant-level lines of function $s_3^{(yz)}(y,z)$ and directed so that the region with higher $s_3^{(yz)}(y,z)$ lies to the left when seeing from the positive end of axis $x$.

We focused upon the simple plane-polarized configuration of Eq. (37) because optical currents in this field were recently considered in detail [23]. However, it is not favourable for studying the specific features and roles of the spin and orbital flows. An essential aspect that differentiates it from the common paraxial fields is that here the spin flow, as well as the orbital one, is directed in accord with predominant beam propagation and constitutes a part of the main (longitudinal) energy flow of the beam. In such a situation, there is no physical difference between the spin and orbital momenta; in fact, they are likely to be indistinguishable and only their (algebraic) sum can be observed. In our opinion, this serves an additional argument for the mechanical equivalence between the spin and orbital momenta and testifies that physical manifestations of both contributions should be the same.

Nevertheless, it would be useful to support this idea by the direct analysis of mechanical action of the spin momentum. To search the conditions where the spin momentum represents itself "in pure form", with all its specific properties, we concentrate on an alternative situation when the superposed waves are circularly polarized, that is, instead of Eq. (37),

$$E_{x1} = \frac{E_0}{\sqrt{2}}, \quad E_{y1} = i\sigma \frac{E_0}{\sqrt{2}}, \quad \sigma = \pm 1. \tag{40}$$

Then an *x*-component of the spin flow appears whose form is dictated by Eqs. (32) and (40):

$$p_{Sx} = -\frac{g}{c}\sigma |E_0|^2 \sin 2\gamma \sin 2\Phi. \tag{41}$$

This *x*-directed spin flow is unique and cannot be contaminated with any orbital contribution; moreover, an *x*-directed energy flow in this geometry seems counter-intuitive and its observation would be an impressive evidence for the physical consistence of the spin momentum.

### 4. Numerical analysis and discussion

To assess the mechanical action of the incident field, we calculated Cartesian components of the force $F_x$, $F_y$, $F_z$ (25) that acts on a probing particle placed within the field satisfying conditions (26), (27) and (40). The results for particles of different physical nature modeled by the relative refraction index *m* are presented in Fig. 2a–c as functions of the diffraction parameter $\xi = ka$ where *a* is the particle radius [24]. Upon calculations, condition

$$2\Phi = -\pi/2 \tag{42}$$

($y = 0$, $\delta = \pi/2$ or $\delta = 0$, $y = \pi/(4k\sin\gamma)$) was chosen that corresponds to maximum absolute value of the spin flow (32) or (41); the angle value $\gamma = 0.01$ rad allows to consider the near-paraxial regime frequently occurring in practice. To decrease the dynamical range of presented data, they are normalized by means of dividing all calculated forces by the total momentum flux of the incident field through the particle cross section,

$$F_0 = 2g\left(|E_{x1}|^2 + |E_{y1}|^2\right) \cdot \pi a^2. \tag{43}$$

Fig. 2a presents results obtained for a metallic reflecting particle (*m* = 0.32+2.65i corresponds to Au particles suspended in water [25]); Fig. 2b was calculated for the model of strongly reflecting dielectric particle (*m* = 200+i), and Fig. 2c describes the behavior of usual dielectric particles (*m* = 1.5 is typical for various glass or latex materials [20]). For comparison, "pure" contributions of the scattered field $F_{ys}$, $F_{zs}$ following from Eq. (22) are also presented by dashed lines ($F_{xs} \equiv F_x$, see the note below Eq. (25)).

All the calculated dependences possess an oscillatory character typical for the Mie scattering [24,25] and explained by resonance properties of the particle (noticeable oscillations in the initial segment of curve $F_z$ in Fig. 2b are an artifact of the non-physically high refraction index). The results of Fig. 2 should be confronted with the incident field momentum components (32) – (34). The longitudinal force (blue curves $F_z$ and $F_{zs}$) represents the usual light pressure effect; the transverse *y*-component (black curves $F_y$ and $F_{ys}$) corresponds to the gradient force that attracts a particle to or repels it from regions of high electromagnetic energy concentration [26]: under condition (42), the energy density (31) possesses the maximum gradient. The gradient nature of the force $F_y$ is confirmed by the linear behavior of solid black curves in Fig. 2a–d at small $\xi$. With account for normalization divider (43), this means that the force is proportional to the particle volume, which is clearly seen in Fig. 3 showing the dependences of Figs. 2c, d at the small-particle region in logarithmic scale.

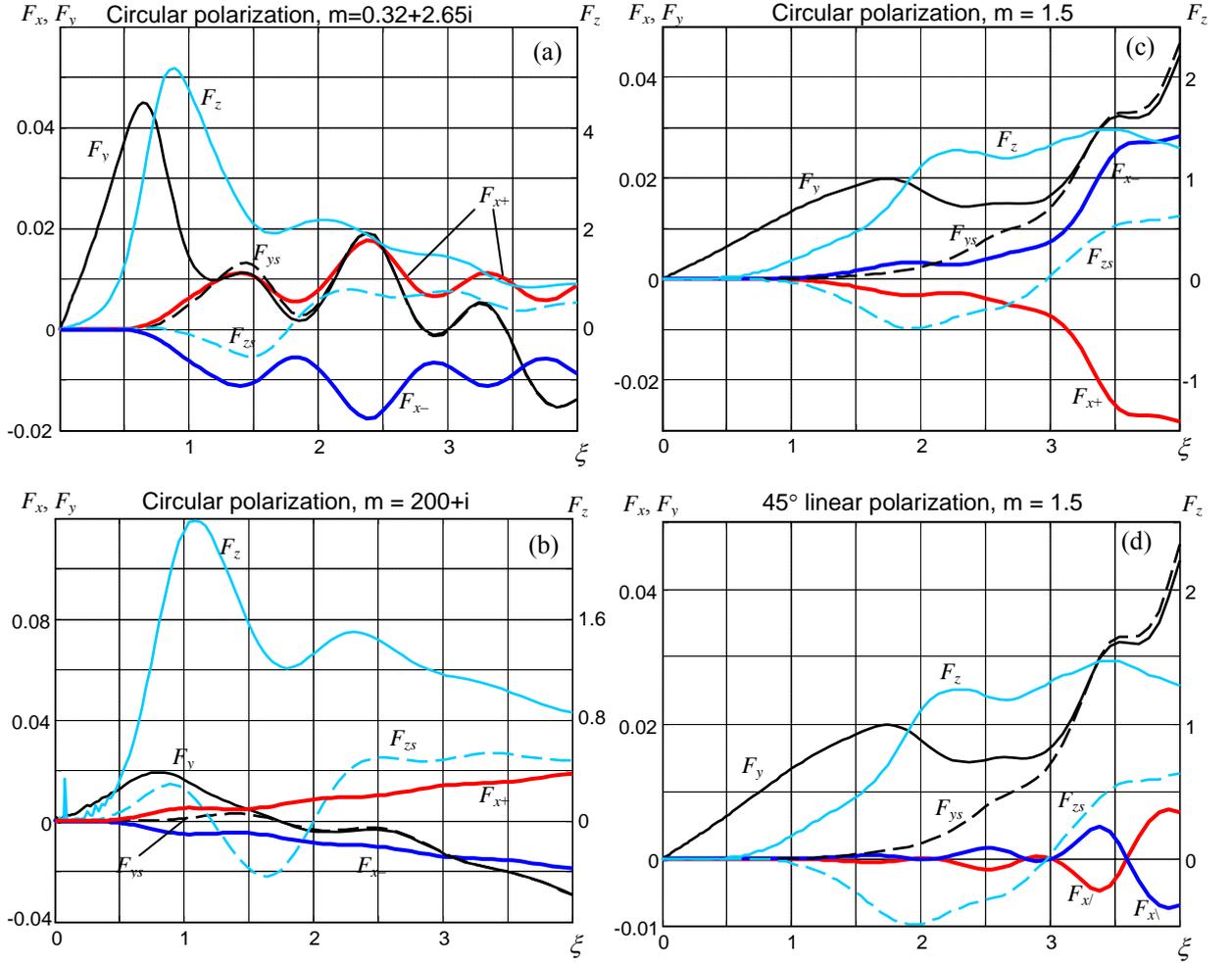

Fig. 2. Components of the force experienced by a spherical particle vs the diffraction parameter $\xi = ka$ in the inhomogeneous incident field described by Eqs. (26) – (34), $\gamma = 0.01$, $\Phi = -\pi/2$, with (a – c) circular polarization of Eqs. (40) and (d) 45° linear polarization of Eqs. (27) ($E_{y1} = \pm E_{x1}$). Data are normalized by $F_0$ (43), curve labels: ($F_y$, $F_z$) Cartesian components of the total force (25), ($F_{ys}$, $F_{zs}$) components of the force (22) (dashed lines); ($F_{x+}$, $F_{x-}$) force component associated with the spin flow (41) for $\sigma = \pm 1$, ($F_{x/}$, $F_{x\backslash}$) x-component of the force in case of 45° linear polarization with $E_{y1} = \pm E_{x1}$. The particle refraction index equals to: (a) $m = 0.32+2.65i$; (b) $m = 200+i$; (c) and (d) $m = 1.5$.

The most interesting results relate to the component $F_x$ – the only force component that can be associated with the spin flow (41). This attributing is directly supported by the fact that, in full agreement with the spin flow behavior, $F_x$ changes the sign with inversion of the incident beam helicity $\sigma$ (compare curves $F_{x+}$ and $F_{x-}$ in Figs. 2a–c) – while all other curves remain the same. Additional arguments for the proposed interpretation follow from analysis of mechanical action of x- or y-polarized incident fields (conditions (37) are fulfilled). Corresponding results are not presented in Fig. 2 because in all cases curves $F_z$, $F_{zs}$, are close to ones calculated for circular polarization but $F_x$ completely vanishes.

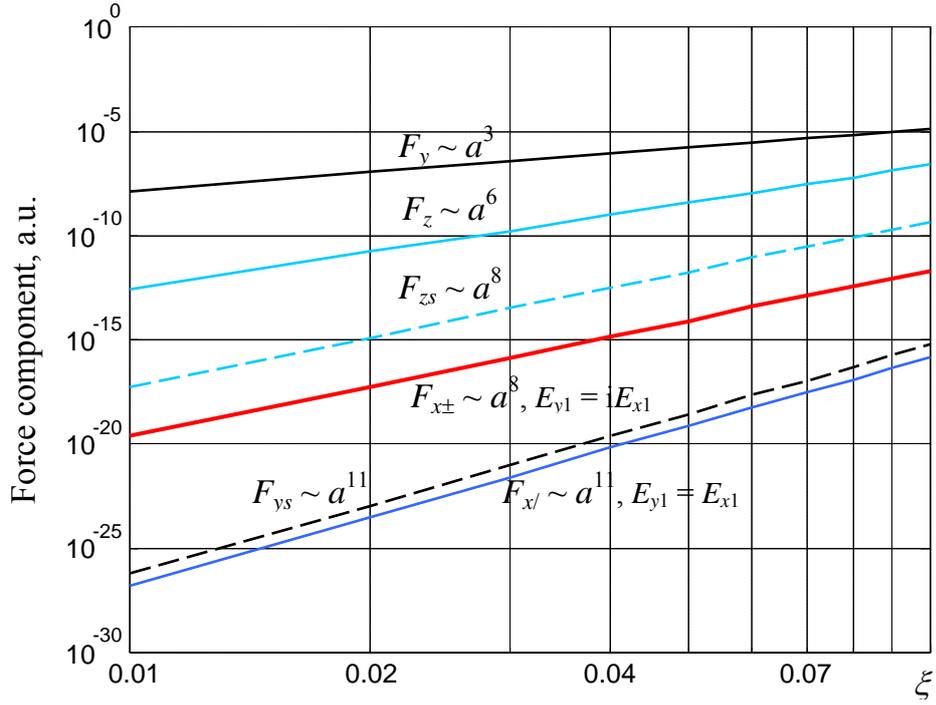

Fig. 3. Particle-size dependence of the field-induced force components acting on a small dielectric particle with refraction index $m = 1.5$ in a circularly polarized ($F_{x\pm}$) and 45° linear polarized ($F_{x/}$) incident field (conditions of Fig. 2c, d) in the double logarithmic scale. Line labels are the same as in Fig. 2 and the orders of growth are added; lines $F_y$, $F_{ys}$, $F_z$ and $F_{zs}$ for both polarization cases are identical.

Similar conclusions can be derived from Fig. 4 that illustrates how the force components depend on the particle position via parameter $\Phi$ (see Eq. (30)). The spatial dependence of $F_y$, $F_{ys}$ exactly reproduces the gradient of the energy density (31) and $F_z$, $F_{zs}$ behave in accordance with the total longitudinal momentum of the incident field (sum of expressions (33) and (34)). $F_x$ varies proportionally to the spin momentum (41) in case of circular polarization (lines $F_{x+}$ and $F_{x-}$ in Fig. 4a) and disappears for the $x$- or $y$- linearly polarized fields (Figs. 4c, d).

All the above arguments witness that the force experienced by a probing particle, really, can reflect the local value of the internal energy flow. However, there are some additional factors that also affect the possible particle's motion and must thus be taken into account in interpretation of the probing particle experiments. The first one is the already mentioned gradient force that is a source of $F_y$ in the considered field configuration. Secondly, Fig. 2c shows that a probing particle can experience the mechanical action directed against the spin momentum (in this panel, signs of $F_{x+}$ and $F_{x-}$ are just opposite to those in Figs. 2a, b and to what is dictated by Eq. (41)). In contrast to known reports of the reverse mechanical action [27], the present effect is expected to take place for spherical particles in homogeneous media. This fact deserves special investigation; at the moment, we may suppose that it is caused by specific character of the particle-induced field distortion together with particular features of the particle-field interaction. Anyway, the spin-induced nature of the force $F_x$ in conditions of Fig. 2c is doubtless for it reverses with changing the sign of $\sigma$ and disappears for $x$- and $y$-polarized incident fields (Fig. 4c, d).

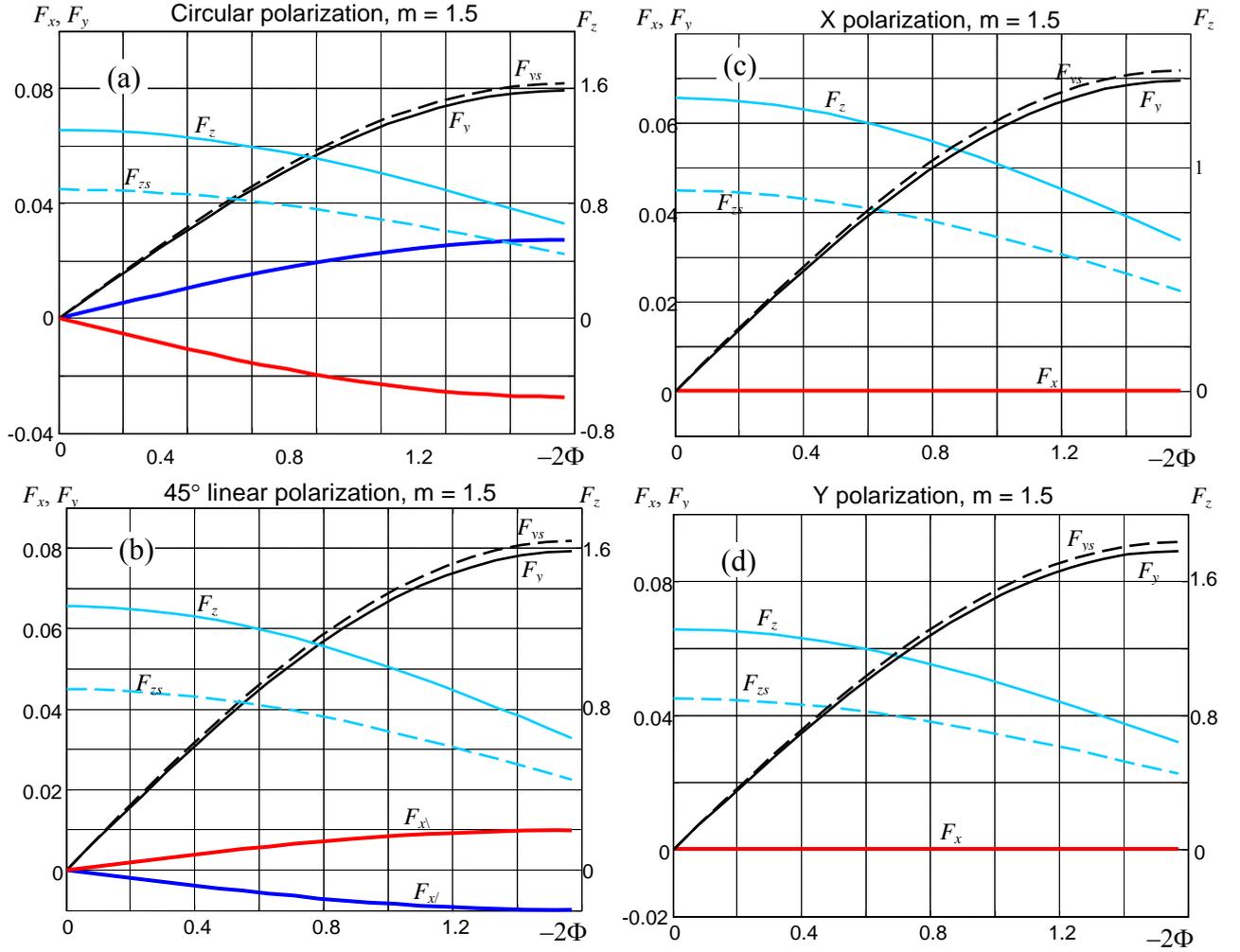

Fig. 4. Coordinate dependence of the force applied to the spherical particle with diffraction parameter $\xi = 5$ and refraction index $m = 1.5$ in the incident field of Eqs. (26) – (34), $\gamma = 0.01$ (conditions of Fig. 2c, d) with: (a) circular polarization of Eqs. (40); (b) 45° linear polarization of Eqs. (27) ($E_{y1} = \pm E_{x1}$); (c) $x$- and (d) $y$-polarizations of Eqs. (37). Force values are normalized by (43).

## 5. Non-Poynting sources of the mechanical action

An important consequence of the performed analysis is the conclusion that an optical field may exert the mechanical force that cannot be associated with any component of the field momentum (energy flow) in the incident beam. A characteristic example is provided by the force $F_y$ in Figs. 2 – 4 that does not vanish despite that under conditions (26), (27) $p_{Sy} = p_{Oy} = 0$. This force does not depend on the polarization state and its spatial variation fairly corresponds to the gradient of the energy density distribution (31) (see Fig. 4a, c, d), so it was identified in the above Section with the known gradient force [20] that pushes a particle to or from the high-intensity regions.

However, our results show that a certain force may appear even in the direction where both the incident field momentum and the energy density gradient equal to zero. For example, an $x$-directed force exists in a linearly polarized incident field provided that, in contrast to conditions (37), both $x$- and $y$-components are non-zero (lines $F_{x/}$ and $F_{x\backslash}$ in Figs. 2d and 4b). This phenomenon can be ascribed to the dipole force that emerges in inhomogeneously polarized fields [10,26]. In paraxial case and for weak electric polarization of the medium exposed to the field, the transverse component of this force exerted to the unit volume of the medium is proportional to [10]

$$\mathbf{F}^{dip} = \frac{\varepsilon_r - 1}{2c}\left[\left(-\mathbf{e}_x\frac{\partial}{\partial x} + \mathbf{e}_y\frac{\partial}{\partial y}\right)s_1 - \left(\mathbf{e}_x\frac{\partial}{\partial y} + \mathbf{e}_y\frac{\partial}{\partial x}\right)s_2\right] \quad (44)$$

where $\varepsilon_r$ is the real part of the medium permittivity, $s_1(x,y)$ and $s_2(x,y)$ are transverse distributions of the first and second Stokes parameters [24]. In the situation of (26), (27) they can be represented similarly to Eq. (36):

$$s_1 = gc\left(|E_x|^2 - |E_y|^2\right) = 2gc\left(|E_{x1}|^2 - |E_{y1}|^2 \cos^2\gamma\right)(1 + \cos 2\Phi), \quad (45)$$

$$s_2 = gc\left(E_x E_y^* + E_y E_x^*\right) = 2gc\left(E_{x1} E_{y1}^* + E_{y1} E_{x1}^*\right)\cos\gamma\,(1 + \cos 2\Phi). \quad (46)$$

Again, like in the above-considered situations, the exact value of the force is mediated by a number of complicated details of the field-particle interaction, so the presence of force (44) can be detected via specific features of its dependence on the particle position and on the incident field polarization. Since quantities (45), (46) depend only on $y$, the general expression (44) reduces to

$$F_x^{dip} \propto gk\left(E_{x1} E_{y1}^* + E_{y1} E_{x1}^*\right)\sin 2\gamma \sin 2\Phi, \quad (47)$$

$$F_y^{dip} \propto -2gk\left(|E_{x1}|^2 - |E_{y1}|^2 \cos^2\gamma\right)\sin\gamma \sin 2\Phi. \quad (48)$$

Hence, the $x$-component (47) vanishes in the $x$- or $y$-polarized field but differs from zero for any other case of plane polarization, and the sign reversal of the product $E_{x1} E_{y1}$ causes its inversion. These features, equally specific for the $x$-component of force, appearing in case of "oblique" linear polarization (curves $F_{x/}$ and $F_{x\backslash}$ in Fig. 2d), allow the latter to be identified with the $x$-component of the dipole force (44). Its spatial dependence (curves $F_{x/}$ and $F_{x\backslash}$ in Figs. 4b) also agrees with Eq. (47) (and, by the way, is identical to that of the spin flow (41) (see Fig. 4a)).

Generally, in the situation of Eqs. (26), (27), the discussed $x$-component of the dipole force (47) acts similarly to the spin flow (41). However, it is associated with linear polarization and vanishes in circularly polarized fields while the force originating from the spin flow, quite oppositely, vanishes in plane-polarized fields. Both forces show different behavior in respect to the particle size: for small particles, the spin-induced action grows proportionally to $a^8$ while the dipole-force action grows as $a^{11}$ (see lines $F_{x\pm}$ and $F_{x/}$ in Fig. 3). The apparent deviation from the particle-volume proportionality dictated by (44) and (47), (48) can be ascribed to approximate character of Eq. (44) derived for the Drude model of weakly polarized media [10] and to details of the field-particle interaction that are omitted in the above estimates (after all, the real dipole force is determined by Eq. (44) with the field parameters measured inside the particle whereas in (47) and (48), characteristics of the unperturbed incident field were used). In the middle-size region, force $F_{x\pm}$ shows the tendency to unidirectional growth, though in oscillatory manner (curves $F_{x+}$ and $F_{x-}$ in Figs. 2a–c), whereas $F_{x/}$ and $F_{x\backslash}$ in Fig. 2d oscillate near the zero line. In real situations, the spin-flow force and the $x$-component of the dipole force seem to act jointly, but actual value and even the sign of each contribution is a complicated function the field characteristics and of the particle optical properties. Detailed study of this problem is out of scope of the present paper.

The $y$-component of the dipole force is less interesting because, according to Eqs. (48) and (31), it acts similarly to the gradient force in all cases. Regarding the field and particle characteristics, it can slightly modify the resulting value of $F_y$, which explains small difference between curves $F_y$ in Figs. 2c, d and 4c, d.

Most impressively, the dipole force action manifests itself in conditions where all other sources of the field mechanical action are absent. This is realized if, instead of (27), the following conditions hold:

$$E_{x1} = E_{x2}e^{-i\delta} = \frac{E_0}{\sqrt{2}}, \quad E_{y1} = -E_{y2}e^{-i\delta} = i\frac{E_0}{\sqrt{2}} \quad (49)$$

(both plane waves of the superposition (9) are circularly polarized but, contrary to (40), their helicities are opposite). Then, taking into account condition (26), from Eqs. (10) – (16) one obtains that $D(y,z) = 0$, $w = 2|E_0|^2 = $ const, and the transverse flow components $p_{Sx} = p_{Sy} = p_{Ox} = p_{Oy} = 0$. However, the dipole force (44) still exists and, due to (47) and (48)

$$F_x \propto 2gk|E_0|^2 \sin\gamma \cos\gamma \cos 2\Phi, \quad F_y \propto -gk|E_0|^2 (1+\cos^2\gamma)\sin\gamma \sin 2\Phi. \quad (50)$$

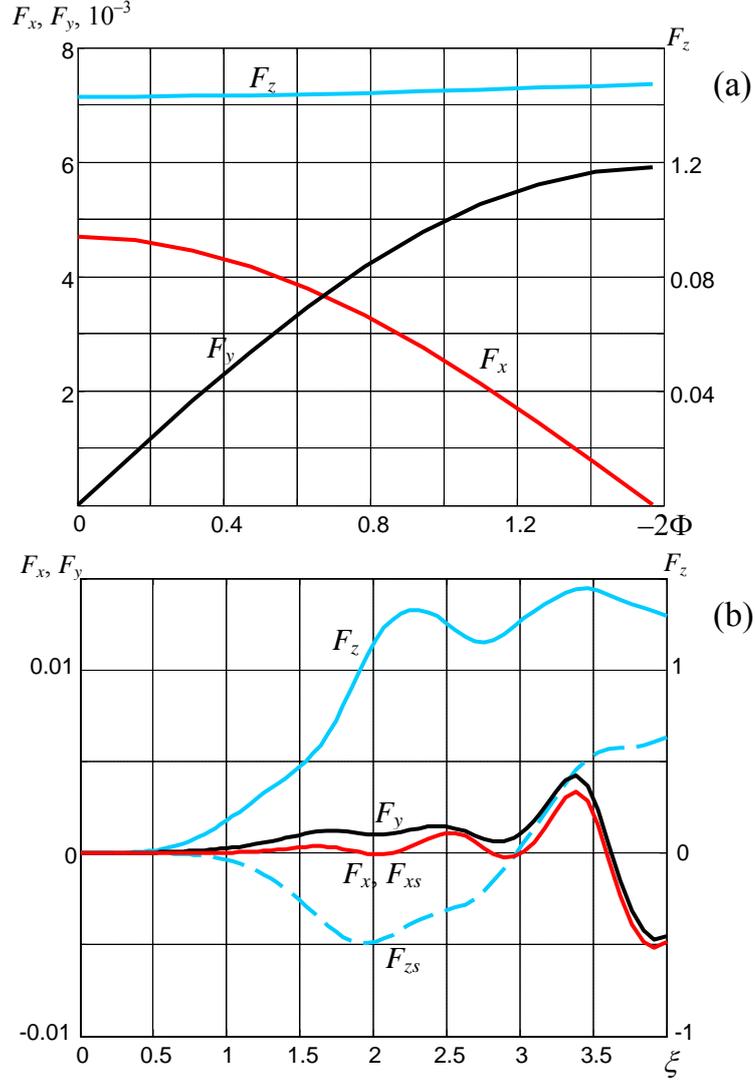

Fig. 5. (a) Coordinate and (b) size dependence of the field-induced force components exerted to the spherical particle with refraction index $m = 1.5$ in the incident field of Eqs. (49) (superposition of plane waves with opposite circular polarizations), $\gamma = 0.01$. In panel (a) the diffraction parameter is $\xi = 3.4$, in (b) the particle position corresponds to $2\Phi = -\pi/4$. The force values are normalized by (43).

Results of the corresponding numerical calculations are illustrated by Fig. 5. Normalized force components $F_x$ and $F_y$ (Fig. 5a) behave in full accordance to Eqs. (50), which confirms that the calculated mechanical action, indeed, can be identified with the dipole force. Even their values are close to each other, as it ought to be from Eqs. (50) at $2\Phi = -\pi/4$. Note that, in full compliance with the homogeneous intensity of the field obeying Eqs. (26), (49), the longitudinal force $F_z$, as a function of $\Phi$, is almost constant (Fig. 5a); its dependence on the particle size parameter $\xi$ is practically the same for different polarization states (compare curves $F_z$, $F_{zs}$ in Figs. 2c, 2d and 5b). Such behavior of the longitudinal force is quite expectable; in contrast, the very existence of the transverse force components seems, at first glance, counter-intuitive, until the dipole force (44) and (50) is employed. The incident field configuration of Eqs. (26) and (49) represents an instructive example of the optical field where the transverse ponderomotive action appears exclusively due to the inhomogeneous polarization, without any intensity gradients and internal energy flows.

**Conclusion**

A model of spatially inhomogeneous optical field is proposed that is formed by superposition of two plane waves. Despite its simplicity, the model adequately represents general properties of inhomogeneous fields, including regularities in behavior of internal energy flow and its spin and orbital parts. Using the Mie theory, the mechanical force acting on a probing particle is calculated numerically. This force consists of several contributions that can be associated with the spin and orbital parts of the internal energy flow of the incident field. In general, this testifies to the mechanical equivalence of the spin and orbital parts of the electromagnetic momentum and shows possibility of detecting the internal flows by translational and/or orbital motion of probing particles suspended within the field.

However, some precautions should be kept in mind when interpreting the probing particle behavior. First of all, the field mechanical action depends on the particle size and refraction index in rather complicated and non-monotonous way (see, e.g., Fig. 2a–d). Besides, the particle strongly disturbs the incident field pattern and in some cases the resulting force is rather far from naïve expectations that the mechanical action is proportional to the local value of the incident field momentum density in the point where the particle is placed. In essence, only the line along which the particle is "moved" by a certain flow component usually coincides with the flow line; the actual force magnitude and even its sign cannot be predicted from the flow pattern alone (e.g., forces $F_{x+}$ and $F_{x-}$ in Fig. 2c are opposite to the corresponding spin momentum (41)).

Second, in inhomogeneous optical fields, forces of other origin, which are not related to the energy flows, may appear whose action can substantially modify and even mask the field momentum action. The first "masking" effect appears due to the gradient force that attracts a particle to or repels it from regions with high energy density; the second one originates from the dipole force (44) that non-trivially depends on the polarization inhomogeneity. Their masking influence can be reduced or eliminated in certain specially designed field configurations. For example, in the geometric arrangement discussed in this paper (Fig. 1), the gradient force "pushes" a particle along the $y$-direction while the internal flows are expected to produce motions along axes $z$ and $x$. The dipole force that, in the considered arrangement, acts "together" with the spin flow, can be separated due to specific relation to the field inhomogeneity: it vanishes for the circular polarization, when the spin flow is maximal, and reaches the maximum at 45° linear polarization while the spin flow completely disappears. The specific feature of the dipole force is that it owes to the polarization inhomogeneity rather than to the intensity inhomogeneity and may appear in situations where the incident field configuration is characterized by homogeneous energy distribution and contains no transverse energy flows.

Basing on the presented examples, we suppose that various situations where an optical field exerts the polarization-dependent mechanical action on isotropic particles (and, of course, the scattered field acquires the corresponding recoil momentum and the polarization-induced spatial anisotropy) can be treated as a new form of the spin-orbit interaction of light. Its origination is likely the same as reported previously for the scattering of a circularly polarized plane wave [28] (generation of optical vortices in the scattered field) but now this phenomenon is accompanied by the symmetry-breaking interference of the scattered fields from different plane waves of the superposition. In fact, the spatial distribution of the scattered field demonstrates some distinct polarization-dependent features that constitute the separate interest and will be considered elsewhere. Here, we only mention that an incident field with the spin momentum induces the scattering anisotropy even in the Rayleigh scattering regime (corresponding corrections to Eqs. (19) are antisymmetric with respect to the longitudinal dimension $z$). The field scattered into the forward hemisphere acquires the integrated transverse momentum; approximately the same momentum but with opposite sign is imparted to the back-scattered field. These values are about two orders of magnitude higher than the total momentum imparted to the particle.

Finally, we emphasize that the model of inhomogeneous optical field developed in this paper, even in its simplest version, provides consistent conclusions that will be useful in planning and performing the probing particle experiments. Besides, the presented model can be easily generalized to describe more complicated situations to reflect fine features of the real optical fields.